\documentclass[reprint, floatfix, aps, pre]{revtex4-1}
\usepackage{amsmath,amssymb}
\usepackage{graphicx,hyperref,color}
\newcommand{\nn}{\nonumber}
\begin{document}
\title{Classical density functional theory for a two-dimensional isotropic ferrogel model with labeled particles}
\author{Segun Goh}
\email{segun.goh@hhu.de}
\affiliation{Institut f\"ur Theoretische Physik II: Weiche Materie, Heinrich-Heine-Universit\"at D\"usseldorf, D-40225 D\"usseldorf, Germany}

\author{Ren{\'e} Wittmann}
\affiliation{Institut f\"ur Theoretische Physik II: Weiche Materie, Heinrich-Heine-Universit\"at D\"usseldorf, D-40225 D\"usseldorf, Germany}

\author{Andreas M. Menzel}
\affiliation{Institut f\"ur Theoretische Physik II: Weiche Materie, Heinrich-Heine-Universit\"at D\"usseldorf, D-40225 D\"usseldorf, Germany}

\author{Hartmut L{\"o}wen}
\email{hlowen@thphy.uni-duesseldorf.de}
\affiliation{Institut f\"ur Theoretische Physik II: Weiche Materie, Heinrich-Heine-Universit\"at D\"usseldorf, D-40225 D\"usseldorf, Germany}
\date{\today}
\begin{abstract}
In this study, we formulate a density functional theory (DFT) for systems of labeled particles,
considering a two-dimensional bead-spring lattice with a magnetic dipole on every bead as a model for ferrogels.
On the one hand, DFT has been widely studied to investigate fluid-like states of materials, 
in which constituent particles are not labeled as they can exchange their positions without energy cost. 
On the other hand, in ferrogels consisting of magnetic particles embedded in elastic polymer matrices,
the particles are labeled by their positions as their neighbors do not change over time. 
We resolve such an issue of particle labeling,
introducing a mapping of the elastic interaction mediated by the springs 
onto a pairwise additive interaction (pseudo-springs) between unlabeled particles.
We further investigate magnetostriction and changes in the elastic constants under altered magnetic interactions 
employing the pseudo-spring potential.
It is revealed that there are two different response scenarios 
in the mechanical properties of the dipole-spring systems: while systems at low packing fractions 
are hardened as the magnetic moments increase in magnitude, 
at high packing fractions softening due to diminishing effects from the steric force,
associated with increases in the volume, is observed.
Validity of the theory is also verified by Monte-Carlo simulations with both real and pseudo-springs.
We expect that our DFT approach may shed light on 
an understanding of materials with particle inclusions. 
\end{abstract}
\maketitle
%
\section{Introduction}
\label{sec:intro}
%
In statistical mechanics, \emph{indistinguishability} of particles and consequently the correct Boltzmann counting
play an essential role, see, e.g., Refs.~\cite{Pathria1996, Kardar2007, Huang2009}. 
The ``Gibbs paradox'' is well-known in this regard and the extensivity of entropy 
is recovered by introduction of the $1/N!$ factor,
which corrects the number of microstates by the number of permutations of $N$ particles.
Commonly, the $1/N!$ factor is regarded as a remnant of quantum mechanics in the classical limit,
in which identical particles are inherently indistinguishable.
In contrast to such a point of view, however, the $1/N!$ factor should be consistently interpreted
based on an ``informatic'' definition of entropy~\cite{Cates2015}. 
Accordingly, 
a modified term, ``undistinguished'' particles~\cite{Sethna2006}, has also been proposed.
Consequently, even though the classical particles such as colloidal particles
are undoubtedly distinguishable, the statistical mechanics with the $1/N!$ correction 
describes the macroscopic behaviors of such systems successfully~\cite{Swendsen2008, Frenkel2014},
as far as one ignores detailed differences between particles~\cite{Jaynes1957, Jaynes1992} 
and leaves the particles unlabeled.

At the very microscopic level, i.e., at the atomic scale, a system consisting of identical particles 
is invariant under permutations of the particles and the particles are unlabeled in principle.
If one considers a mesoscopic length scale 
and employs a coarse-grained description~\cite{Doi1986,Menzel2019}, however, 
it may become necessary to distinguish between particles that are physically identical. 
Such a scenario can emerge if the particles are permanently localized with respect to their neighbors, 
thereby rendering the system non-ergodic on the relevant energy scale.
It is thus a major challenge for statistical mechanical theories 
to describe a model with labeled particles~\cite{Cremer2017} 
or to keep track of a single localized particle~\cite{Wittmann2019}. 

Here we develop a statistical description for
a new class of composite materials,
which consist of magnetic particles and an elastic polymer matrix~\cite{Filipcsei2007, Ilg2013, Menzel2019}.
In these materials, called ferrogels or magnetorheological elastomers, 
the dynamical trajectories of the magnetic particles are frequently strongly constrained 
by the polymeric environment~\cite{Gundermann2014, Landers2015}.
Such a magneto-mechanical coupling can even be enhanced by directly anchoring the polymers 
on the surface of magnetic particles~\cite{Frickel2011, Messing2011, Ilg2013, Roeder2015}.
Hence, the elastic properties of the materials can be tuned from outside
by non-invasive applications of magnetic fields~\cite{Stepanov2008, Stolbov2011, Stolbov2019}.
As a further consequence of this coupling the particles cannot exchange their positions 
due to the fixation by the elastic medium.

Various studies have been conducted to theoretically understand the behavior of such ferrogels 
with different description levels from the microscopic scale resolving the individual polymer particles
to the macroscopic hydrodynamic/thermodynamic theory. 
For many practical purposes, one may neglect the thermal motions of the magnetic particles~\cite{Menzel2019}.
In particular, a mesoscopic dipole-spring model has been adopted to study the elastic and dynamical 
properties of ferrogels~\cite{Pessot2016, Pessot2018, Goh2018}. 
The matrix-mediated interaction between magnetic particle inclusions
has also been revealed in terms of continuous elastic backgrounds~\cite{Biller2014, Biller2015, Cremer2015, Cremer2016, Puljiz2016, Puljiz2017}.
Furthermore, microscopic descriptions of ferrogels via coarse-grained molecular dynamics simulations
enable us to probe the role of thermal motions of the magnetic particles explicitly~\cite{Weeber2012, Weeber2015jmmm, Weeber2019}.

Here, we merge several of the aspects mentioned above.
Our goal is to formulate a statistical mechanical theory for ferrogels 
in a dipole-spring model with thermal fluctuations taken into account.
The most challenging problem that has to be addressed along the way arises from the fact that the
particles in ferrogels are strictly \emph{labeled} by their positions as in lattice systems, for instance,
the classical Ising or XY models~\cite{Goldenfeld1992} and harmonic crystals~\cite{Ashcroft1976}. 
Accordingly, correcting the number of microstates by the factor $1/N!$ does not apply to 
a statistical description of ferrogels and a permutation of particles will cause a change in energy
(physically this results in strong distortions of the surrounding elastic matrix).
While computational approaches such as Monte-Carlo (MC) simulations are still feasible~\cite{Weeber2019}, 
the formulation of a statistical mechanical theory is severely complicated by
the inherent composite nature of the ferrogels, in contrast to, for instance, harmonic crystals.
In practice, one would need 
to take into account the nonlinearity stemming from the steric and magnetic interactions.

One natural candidate for a statistical theory is classical
density functional theory (DFT)~\cite{Evans1979, Lutsko2010, Evans2016}, 
which has been probed to be successful for variety of systems~\cite{Lowen2002}, 
ranging from simple classical fluids~\cite{Ebner1976} 
to systems showing a freezing transition~\cite{Ramakrishnan1979},
from hard-spheres~\cite{Roth2010} to hard convex particles~\cite{Wittmann2016}  
and also for two-dimensional systems~\cite{Roth2012, Wittmann2017, Lin2018},
including dipolar or electrostatic interactions~\cite{Zimmermann2016, Roth2016}, and capturing
the spinodal decomposition dynamics~\cite{Archer2004} in an adiabatic approximation for time-dependent systems.
However, due to the particle labeling, 
a direct application of the machinery of DFT to ferrogels is not possible.
To this end, we will map the elastic interaction onto an appropriate pairwise pseudo-potential~\cite{Cremer2017}
between unlabeled particles, which allows us to formulate a DFT.
Ultimately, we aim at investigating the elastic properties of the dipole-spring systems within the DFT framework.
By comparison to the MC simulations, the validity of the theory is confirmed.

The paper is organized as follows: In Sec.~\ref{sec:model}, we introduce a two-dimensional model for ferrogels.
Sec.~\ref{sec:mapping} describes detailed procedures of the mapping and the subsequent formulation of 
a DFT. Combining MC simulations and DFT calculations, 
two distinctive response scenarios in elastic properties 
to the change of the magnetic moment are identified in Sec.~\ref{sec:elastic}. 
Lastly, a summary and a discussion are given in Sec.~\ref{sec:conclusion}.



%
\section{The dipole-spring model}
\label{sec:model}
\begin{figure}
\includegraphics[width=8.6cm]{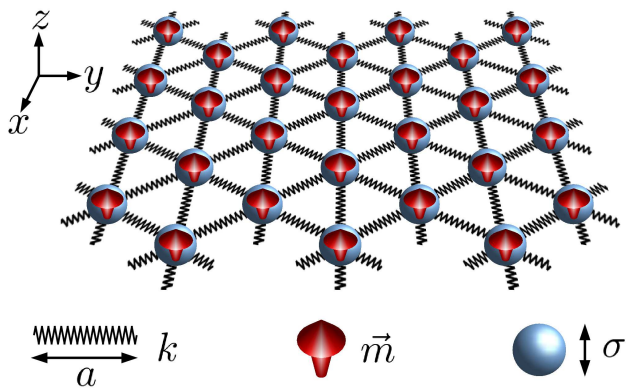}
\caption{\label{fig:model} Illustration of the dipole-spring model.
}
\end{figure}

We consider a bead-spring model~\cite{Doi1986} in terms of  a periodic two-dimensional hexagonal lattice, 
as illustrated in Fig.~\ref{fig:model}.
There are $N$ identical magnetic particles and $3N$ identical harmonic springs connecting the nearest neighbors.
We denote the position and the dipole moment of the $i$th particle of diameter $\sigma$ by $\vec{r}_i$ and $\vec{m}_i$, respectively.
The total Hamiltonian $\mathcal{H}_{\rm tot}$ of the model system is given by the sum of the kinetic part
and the interaction Hamiltonian $\mathcal{H}_{\rm int}$ 
which consists of three parts in the form of
\begin{align}
\mathcal{H}_{\rm int} = \mathcal{H}_{\rm m} + \mathcal{H}_{\rm el} + \mathcal{H}_{\rm st}.
\label{eq_Hint}
\end{align}
Among these three terms, the magnetic part $\mathcal{H}_{\rm m}$ and the steric part $\mathcal{H}_{\rm st}$, 
can be written as
\begin{align}\label{eq:pairadd}
\mathcal{H}_{\rm m, st} = \frac{1}{2} \sum_{i\neq j} u_{\rm m, st} (\vec{r}_{ij}),
\end{align}
where $\vec{r}_{ij} = \vec{r}_j - \vec{r}_i$.
For isotropic interactions, the vector $\vec{r}_{ij}$ in the argument 
can be replaced by $r_{ij} = |\vec{r}_{ij}|$.

First, the two-body magnetic dipole-dipole interaction energy between magnetic particles reads
\begin{align}
u_{\rm m}(\vec{r}_{ij}) = \frac{\mu_0}{4\pi} 
\left[ \frac{\vec{m}_i\cdot \vec{m}_j}{r_{ij}^3}-
\frac{3(\vec{m}_i\cdot \vec{r}_{ij})(\vec{m}_j\cdot \vec{r}_{ij})}{r_{ij}^5} \right],
\end{align}
where $\mu_0$ is the vacuum permeability. 
Henceforth, we assume that the magnetic moment is constant in time and for all particles,
i.e., $\vec{m}_i (t)=\vec{m}$ regardless of $i$ and $t$ for the sake of simplicity.
Moreover, we constrain ourselves to isotropic magnetic interactions in the two-dimensional plane
by further assuming 
that $\vec{m} = m\hat{z}$ is perpendicular to the lattice plane, 
so that the magnetic two-body interaction energy can be written in a simpler form as
\begin{align}
u_{\rm m}(r_{ij}) = \frac{\mu_0 m^2}{4\pi} \frac{1}{r_{ij}^3}.
\label{eq_Hmag}
\end{align}

The second term in Eq.~\eqref{eq_Hint} corresponds to the elastic energy of the harmonic springs 
and reads
\begin{align}
\mathcal{H}_{\rm el} = \sum_{\langle i, j \rangle} u_{\rm el}(r_{ij}) =\sum_{\langle i, j \rangle} \frac{1}{2} k_{\rm el}({r}_{ij} -a)^2,
\label{eq_Hel}
\end{align}
where $k_{\rm el}$ is the spring constant and $a$ is the rest length of the springs.
Apparently, each spring connects a certain prescribed pair of particles,
and our summation persistently runs only over nearest-neighboring pairs 
as indicated by the angular bracket. 
Consequently, all particles are labeled by the predetermined ordering on a lattice,
the energetic memory of which being recalled by $\mathcal{H}_{\rm el}$.
We remark that Eq.~\eqref{eq_Hel} cannot be cast into the form of
$\frac{1}{2}\sum_{i\neq j} \bar{u}_{\rm el} (\vec{r}_{ij})$ with a $i, j$-independent function $\bar{u}_{\rm el} (\vec{r})$.

Lastly, the two-body steric repulsion energy is taken as a hard-core potential of the form
\begin{equation}
\label{eq_Hst}
u_{\rm st}(r_{ij}) =\begin{cases}
	0 & \ {\rm if}  \ r_{ij} \geq \sigma \\
	\infty & \ {\rm otherwise}
\end{cases}
\end{equation}
and completes the interaction Hamiltonian in Eq.~\eqref{eq_Hint}.
This term prevents the possible divergence of the magnetic interactions at $r_{ij} = 0$.
Following Refs.~\cite{Bernard2009,Roth2012,Engel2013,Lin2018}, 
we introduce a dimensionless packing fraction defined as the ratio of the space occupied by particles to 
the two-dimensional ``volume'' (the term ``volume'' is used for area throughout the paper) of the system. 
As we might consider the systems under constant pressure
(see Sec.~\ref{sec:impl}), however, the volume $V$ of our system is not necessarily a fixed variable.
We therefore define a reference volume $V_{\rm ref} \equiv N V_0 \equiv N \sqrt{3}a^2/2$
in which the springs are in their rest state.
Accordingly, the packing fraction of a reference system with this volume is given 
by $\eta_0=(\sigma^2 \pi/2)/(\sqrt{3} a^2)$.
Apart from $a$, which will be used as the unit of length,
$\eta_0$ depends only on the diameter of particles $\sigma$ and therefore 
we employ it as a model parameter representing the steric repulsion.
In contrast to that, the conventional packing fraction $\eta \equiv (\pi \sigma^2 /4)N/V$ defined in terms of
the ``actual'' volume $V$ of the system, may change as the volume increases/decreases in response 
to a decrease/increase of the pressure or an increase/decrease of the magnetic moment $m$ at constant pressure.

From now on, we measure lengths and energies in units of $a$ and $k_B T$, respectively.
Accordingly, the magnetic moment $m$ and the spring constant $k_{\rm el}$ are measured 
in units of $m_0 \equiv \sqrt{k_B T a^3 /\mu_0}$ and $k_B T/a^2$, respectively.
Finally, let us note that our bead-spring model is identical to the classical harmonic crystal~\cite{Ashcroft1976},
if the magnetic interactions energy $\mathcal{H}_{\rm m}$
and the steric repulsion energy $\mathcal{H}_{\rm st}$ are neglected in Eq.~\eqref{eq_Hint}.
%
\section{Mapping onto pseudo-springs}
\label{sec:mapping}
%
Now we address the issue of particle labeling and 
derive an approximate elastic Hamiltonian $\tilde{\mathcal{H}}_{\rm el}$, 
which can be readily used for the density functional calculation.
Putting aside the magnetic and steric parts of the Hamiltonian, i.e., $\mathcal{H}_{\rm m, st}$, 
which are free from this issue, see Eq.~\eqref{eq:pairadd}, 
we only consider the elastic part $\mathcal{H}_{\rm el}$ in this section.
Following Ref.~\cite{Cremer2017}, we consider a mapping of $\mathcal{H}_{\rm el}$ 
onto a pseudo-spring potential between unlabeled particles of the form
\begin{align}
\mathcal{H}_{\rm el} \to \tilde{\mathcal{H}}_{\rm el} =\frac{1}{2} \sum_{i \neq j} u_{\rm pel}(r_{ij}),
\label{eq_HelMAP}
\end{align}
where $u_{\rm pel}(r)$ is a two-body interaction between particles of center-to-center distance $r$. 
$\tilde{\mathcal{H}}_{\rm el}$, in contrast to Eq.~\eqref{eq_Hel}, involves all particle pairs $(i,j)$. 
Consequently, the Hamiltonian $\tilde{\mathcal{H}}_{\rm el}$ is invariant 
under permutations of the particles, i.e., 
\begin{align} \label{eq:permutation}
\tilde{\mathcal{H}}_{\rm el}(\{ \vec{r}_i \}) = \tilde{\mathcal{H}}_{\rm el}(\{\vec{r}_{\hat{\pi}({i})}\}),
\end{align}
where $\hat{\pi}$ is a permutation operator constituting the symmetric group $S_N$. 
While such a mapping is convenient from a technical point of view, 
it is physically appropriate only if we can ensure through the form of $u_{\rm pel}(r)$ 
that each particle in effect only interacts with a prescribed set 
of other particles, i.e., the same number of nearest neighbors as in the real-spring system.
To this end, we will cut each spring-like interaction (which we specify later) at larger distances 
to prevent interactions between particles too far away from each other. 
The crucial point is then to ensure that $\tilde{\mathcal{H}}_{\rm el}$ does neither introduce additional contacts 
nor miss actually existing ones.

In one spatial dimension, the particles described by a pseudo-spring model are automatically ``labeled'' 
through the steric hard-core repulsions~\cite{Tonks1936,Percus1976}, which fixes their mutual ordering.
Hence, within a proper range of the spring length at sufficiently high density~\cite{Cremer2017}, 
where the particles usually interact 
with their two nearest neighbors, 
the mapping works perfectly.
Note that there is no phase transition due to the strong thermal fluctuations in one dimension. 
In a more realistic two- or three-dimensional fluid, however, 
the particles can always find a path to bypass each other. 
Still, a similar fixation to a cell surrounded by the nearest neighbors can be achieved 
via ergodicity breaking associated with the freezing transition.
In this sense, we can construct a mapping of the labeled particles in the original lattice model 
onto the crystalline phase of the unlabeled particles with all-to-all pairwise-additive interactions.
We take this viewpoint as the inversion of the following interpretation from Ref.~\cite{Cates2015} for a
system of unlabeled hard spheres.
For hard-sphere crystalline systems, even though there are $N!$ possible distinct crystals of labeled particles
as the Hamiltonian is invariant under permutations, 
all microscopic configurations correspond to one unique macroscopic lattice structure.
In this way, lattice models of labeled particles, 
for which the $1/N!$ factor is omitted in the statistical counting, 
provide a good approximation for the crystalline state of materials, 
emerging from the freezing of fluids with undistinguished particles. 
Inversely, for ferrogels, given the lattice system of labeled particles as the original model,
we restore the $1/N!$ factor by mapping the unique lattice
onto $N!$ possible frozen states of a fluid-like system with unlabeled particles.

In Sec.~\ref{sec:just} 
and in Fig.~\ref{fig:pseudo}, our argument justifying 
such a mapping in two dimensions is laid out in full detail, 
before establishing the specification of the elastic pseudo potential $u_{\rm pel}$
in Sec.~\ref{sec:pair}.
Then we demonstrate in Sec.~\ref{sec:DFT} how to implement the mapped system within our DFT approach. 
Before we proceed, we make two further remarks. First, we assume that 
the pseudo-spring systems governed by $u_{\rm pel}$ have a crystalline phase. 
Second, we consider two-dimensional crystals~\cite{Zahn2000, Keim2004} in this study.
As is well known, there is no true long-range order in two-dimensional systems
in the absence of truly long-range interactions~\cite{Mermin1966}.


\subsection{Detailed justification of the mapping}
\label{sec:just}

Our argument is based on the idea that, at low temperatures, the energetic contribution
overwhelms the entropic contribution involving the $1/N!$ factor, enabling us to map the systems of labeled
particles onto those of the unlabeled ones.
Such an argument is strong enough to justify the mapping, for instance, at zero temperature, for which
the minimum of $\mathcal{H}_{\rm int}$ completely determines the equilibrium properties.
At finite temperatures, however, the elastic properties (or even the stability of the systems)
critically depend on the details 
of the mapping and of the profile of the consequent interaction potential 
$\tilde{\mathcal{H}}_{\rm int} \equiv \mathcal{H}_{\rm m} +\tilde{\mathcal{H}}_{\rm el}+ \mathcal{H}_{\rm st}$.
This does not only apply at the minimum of $\tilde{\mathcal{H}}_{\rm int}$, but even if the adjacent particles 
are not located exactly at their lattice sites due to thermal fluctuations.
Accordingly, a quantitative matching of lattice structures in real- and pseudo-spring systems 
turns out to be essential. Therefore, in this section, we carefully describe 
how the $1/N!$ factor and the fluctuations of particles confined to their lattice sites
in the low-temperature crystalline phase should be addressed throughout the mapping. 

Let us consider the ${4N}$-dimensional \emph{phase space} of our system.
The probability density to locate our system in this phase space reads
$\omega (\{ \vec{r}_i\}, \{\vec{p}_i\}) \propto \exp{\{-\beta \mathcal{H}_{\rm tot}(\{ \vec{r}_i\}, \{\vec{p}_i\})\}}$
with the inverse temperature $\beta \equiv k_B T$.
We then partition the phase space into $N!$ subspaces that are related by permutations of the particles,
analogously to the symmetry-related regions discussed in Ref.~\cite{Goldenfeld1992}. 
To this end, we first specify the completely ordered set (subspace) as 
$\sigma_1 = \{ (\vec{r}_1, \ldots, \vec{r}_N,\vec{p}_1, \ldots, \vec{p}_N)|\  |\vec{r}_i| > |\vec{r}_j| \ {\rm for}\ i > j\}$.
In other words, every configuration in which each particle with a smaller label is located closer 
to the origin than each other particle with a larger label belongs to this subspace.
(If there exist pairs of particles, the distances of which from the origin are the same, one can further order
the particle pairs by comparing the angle, e.g., from the $x$-axis.) 
Then, with the permutation operators $\hat{\pi}_i$ (for $i=1, \ldots, N!$ where $\hat{\pi}_1$ is the identity 
of the $S_N$ symmetric group) introduced in Eq.~\eqref{eq:permutation}, one can generate $N!$ subspaces exhausting the whole phase space by permuting particles, i.e.,
$\sigma_i = \{(\vec{r}_{\hat{\pi}_i (1)}, \ldots,\vec{r}_{\hat{\pi}_i (N)},  \vec{p}_{\hat{\pi}_i (1)},$ $ \ldots, \vec{p}_{\hat{\pi}_i (N)})|(\vec{r}_1, \ldots,\vec{r}_N,\vec{p}_1,\ldots , $ $ \vec{p}_N ) $ $\in \sigma_1\}$.
For example, a subspace $\sigma_\alpha$ is generated by the permutation of particles  
in every element (configuration) of the completely ordered subspace $\sigma_1$, 
with a corresponding operator $\hat{\pi}_\alpha$. 

\begin{figure}
\includegraphics[width=8.6cm]{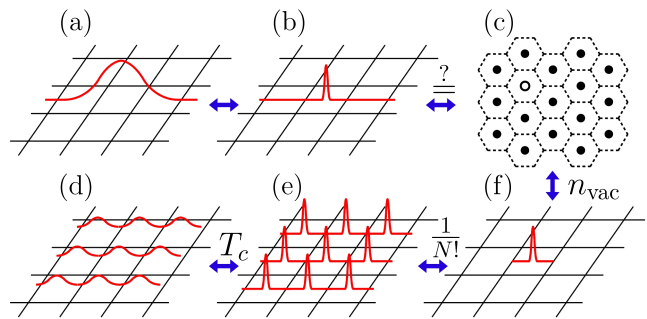}
\caption{\label{fig:pseudo}
A schematic diagram to illustrate the justification of our mapping. In panels (a), (b), and (d)--(f), we present
the whole $4N$-dimensional \emph{phase space} partitioned by the black grids, each cell of the grid representing
a permutation-related subspace $\sigma_i$. Then probability density profiles in the phase space 
are depicted by red lines. 
Specifically, the probability densities of the real-spring systems at high and low temperature are illustrated in panels (a) and (b), respectively; panels (d) and (e) depict those using pseudo-springs at high and low temperature, respectively.
In panel (f), one density profile, localized within one single permutation-related subspace,
is chosen arbitrarily from $N!$ available subspaces [the $1/N!$ factor is simultaneously omitted in the partition function, see the last equality in Eq.~\eqref{eq:Z_pseudo}]. 
Moreover, panel (c) describes the lattice structure in the \emph{configurational space} corresponding 
to panel (f) [see, Eq.~\eqref{eq:Z_pseudo}] or approximately to panel (b) [see, Eq.~\eqref{eq:Z_real}].
Finally, the approximation involved in between panels (b) and (c) is optimized via Eq.~\eqref{eq:cond}, 
which we here achieve by searching for a vanishing concentration $n_{\rm vac}$ of vacancies 
[indicated by an open circle in panel (c)].}
\end{figure}

In the case of the real-spring systems, the probability densities  of each subspace
are not identical to each other, see Figs.~\ref{fig:pseudo}(a) and (b),
as an exchange of any particle pair is always accompanied with a change in energy, i.e., $\mathcal{H}_{\rm el}(\{\vec{r}_i\}) \neq \mathcal{H}_{\rm el}(\{\vec{r}_{\hat{\pi}(i)}\}) $.
However, for pseudo-spring systems, 
the probability densities corresponding to each subspace are identical to each other, 
as illustrated in Figs.~\ref{fig:pseudo}(d) and (e), because the Hamiltonian is invariant under permutations.
Therefore, the probability densities in the whole phase space of the real- and pseudo-spring systems 
are not equivalent to each other in general. More specifically, for real-spring systems, the probability density
in a subspace should deviate from the one in another subspace as they involve permutations of particles, while,
for pseudo-spring systems, they are still identical due to the symmetry under permutations.

Let us now turn to low-temperature systems, i.e., $k_B T \ll k_{\rm el} a^2$.
Recall that there is no phase transition for harmonic crystals~\cite{Jancovici1967, Ashcroft1976},
that is, for our real-spring system. 
Nonetheless, the corresponding probability density becomes highly localized,
as depicted in Fig.~\ref{fig:pseudo}(b).
In the case of the pseudo-spring system, there might occur a freezing transition instead,
which is the starting point of our mapping.
First we constrain ourselves to each one of the permutation-related subspaces. Therefore,
particle exchanges or, namely, permutations of particles via, e.g., vacancy hopping is ignored 
in the following analysis.
Then we consider the ergodicity breaking of the systems due to freezing.
Accordingly, the probability density in each permutation-related subspace is isolated 
as illustrated in Fig.~\ref{fig:pseudo}(e).  
[In contrast to that, as shown in Fig.~\ref{fig:pseudo}(a) and (d), 
the probability densities at high temperature are rather broad for both the real- and pseudo-spring systems
and, therefore, it is not possible to construct such a mapping in our situation.]
In both systems, we conclude that, below a certain temperature,
the thermal vibration in the particle positions are much smaller 
than inter-particle distances in the crystalline state.
Therefore, the trajectory of each particle in each subspace  
does not span the full \emph{configurational space} $\mathbb{R}^2$
but only a localized area. That is, particle $i$ remains localized
within the corresponding Wigner-Seitz cell $\Omega_i$
associated with its average position, see Fig.~\ref{fig:pseudo}(c).
In such a way, we provide a tiling of $\mathbb{R}^2 = \Omega_1 \cup \cdots \cup \Omega_N$, where  
$\Omega_i\cap \Omega_j= \emptyset$ for $i \neq j$.

Now let us be more explicit and consider only one subspace among all permutation-related subspaces 
at low temperature.
In the case of the real-spring system, 
a particle $i$ is localized in the Wigner-Seitz cell 
around a lattice site with the same label $i$,
so that $\int_{\mathbb{R}^2} {\rm d}\vec{r}_i e^{-\beta \mathcal{H}} \approx \int_{\Omega_i} {\rm d}\vec{r}_i e^{-\beta \mathcal{H}}$.
Then, the partition function becomes 
\begin{align}\label{eq:Z_real}
Z_N^{\rm real} &= \frac{1}{\Lambda^N} 
\int_{\mathbb{R}^2} {\rm d}\vec{r}_1 \cdots  \int_{\mathbb{R}^2} {\rm d}\vec{r}_N\, 
e^{-\beta \mathcal{H}_{\rm int}} \nn \\
&\to 
\frac{1}{\Lambda^N}
	\int_{\Omega_1} {\rm d}\vec{r}_1 \cdots  \int_{\Omega_N} {\rm d}\vec{r}_N\, e^{-\beta \mathcal{H}_{\rm int}},
\end{align}
where $\Lambda$ is the mean thermal wavelength of the particles.
As indicated by the arrow on the second line, the partition function has been rewritten 
in terms of the Wigner-Seitz cells. 
With the Wigner-Seitz cells $\{\tilde{\Omega}_i \}$ corresponding 
to the pseudo-spring system in the crystalline state, the partition function involving the unlabeled particles 
can be similarly rewritten as
\begin{align} \label{eq:Z_pseudo}
Z_N^{\rm pseudo} 
&= \frac{1}{\Lambda^N N!} 
\int_{\mathbb{R}^2} {\rm d}\vec{r}_1 \cdots  \int_{\mathbb{R}^2} {\rm d}\vec{r}_N\, e^{-\beta \tilde{\mathcal{H}}_{\rm int}} \nn \\
&\to  \frac{1}{\Lambda^N N!} \sum_{\pi \in S_N} 
	\int_{\tilde{\Omega}_{\hat{\pi}(1)}} {\rm d}\vec{r}_1 \cdots  \int_{\tilde{\Omega}_{\hat{\pi}(N)}} {\rm d}\vec{r}_N\, e^{-\beta \tilde{\mathcal{H}}_{\rm int}} \nn \\
&= \frac{1}{\Lambda^N}
	\int_{\tilde{\Omega}_1} {\rm d}\vec{r}_1 \cdots  \int_{\tilde{\Omega}_N} {\rm d}\vec{r}_N\, e^{-\beta \tilde{\mathcal{H}}_{\rm int}},
\end{align}
where the last equality follows from the fact that $\tilde{\mathcal{H}}_{\rm int}$ is invariant 
under permutations.  
In this way, one can cancel the $N!$ counting factor, completely neglecting the exchange of particles,
as illustrated in Fig.~\ref{fig:pseudo}(f). 
We stress that the probability of the ignored configurations in both Eq.~\eqref{eq:Z_real} and Eq.~\eqref{eq:Z_pseudo} is negligible and vanishes as $T\to 0$ and/or $N\to \infty$.

Comparing now both expressions to each other, 
we map the lattice of labeled particles to that of unlabeled particles via
\begin{align} \label{eq:cond}
\int_{\Omega_1} {\rm d}\vec{r}_1  \cdots \int_{\Omega_N} &{\rm d}\vec{r}_N 
\, e^{-\beta \mathcal{H}_{\rm int}} \nonumber \\
&\stackrel{!}{=} \mathcal{N} \int_{\tilde{\Omega}_1} {\rm d}\vec{r}_1 \cdots \int_{\tilde{\Omega}_N} {\rm d}\vec{r}_N 
\, e^{-\beta \tilde{\mathcal{H}}_{\rm int}},
\end{align}
where $\mathcal{N}$ is an arbitrary constant which does not alter any physical properties of the mapping, indicating that the (many-body) pseudo-spring potential can be determined up to an additive constant.
Even though the above equation is not easy to analyze as it still involves interactions between many particles,
the issue of particle labeling has been resolved: the $1/N!$ factor does not appear in the mapping.
Moreover, it provides us with a route of how to construct an approximate 
elastic part of the Hamiltonian $\tilde{\mathcal{H}}_{\rm el}$ in Eq.~\eqref{eq_HelMAP} for practical calculations.
In addition to the trivial condition that 
(i) the nearest-neighbor interaction of $\mathcal{H}_{\rm int}$ should be reproduced 
by $\tilde{\mathcal{H}}_{\rm int}$, Eq.~\eqref{eq:cond} indicates that 
(ii) the lattice of the real-spring system should be recovered with the pseudo-spring potential, namely,  
$\{\Omega_i \}\stackrel{!}{=}\{\tilde{\Omega}_i \}$.
Next, we describe our strategy to explicitly perform such a mapping by introducing a cut-off for the springs.

\subsection{Pair potential of pseudo-springs}
\label{sec:pair}

%
We consider an elastic interaction energy via pseudo-springs with a cut-off.
The latter introduces an additional degree of freedom to control the offset of the remaining part of the springs, 
leading to a two-body interaction in the form
\begin{equation}
\label{eq:pseudo_spring}
u_{\rm pel}(r_{ij})=\begin{cases}
	\frac{1}{2}k_{\rm el} (r_{ij}-a)^2 - u_{\rm pel}^0, & 
 r_{ij} < R_c  \\
	0, & {\rm otherwise},
\end{cases}
\end{equation}
where $k_{\rm el}$ and $a$ take the same role as in Eq.~\eqref{eq_Hel} 
and the two mapping parameters $R_c$ and $u_{\rm pel}^0$ are fixed by the conditions (i) and (ii), respectively, 
as discussed below.

Regarding condition (i), the nature of the mapping onto the pair potential $u_{\rm pel}(r_{ij})$ 
leads to an apparent violation of Eq.~\eqref{eq:cond}, 
because the number of nearest neighbors cannot be unambiguously imposed. 
Even when the potential has only a finite range,
additional contacts with next-to-nearest-neighbor particles can be formed
in one direction, simultaneously missing the contacts with nearest-neighbor particles in the other directions.
Specifically, if a particle is located, e.g., at one of the corners of its Wigner-Seitz cell,
the minimal possible distance to a particle in a non-adjacent cell is only one edge length, i.e., $a$,
while the maximal possible distance to another particle in a neighboring cell is 
as large as $\sqrt{13}a\approx3.6a$.
This drawback cannot be overcome by any choice of another cell shape but becomes less severe 
as $k_B T/k_{\rm el}a^2 \to 0$.
Hence, we simply consider an isotropic pseudo-potential of range $R_c$ for computational convenience.
This cut-off parameter in Eq.~\eqref{eq:pseudo_spring} should be determined to achieve 
the optimal connectivity with the six nearest-neighbor particles 
(see the inset of Fig.~\ref{fig:dft_formulation}), 
as the best approximation for the original bead-spring Hamiltonian.

To this end, we first determine the cut-off radius $R_c^0$ for the reference system 
with $V_{\rm ref} = N V_0$.
As a simple analytic estimate for $R_c^0$, one may assume a uniform (fluid-like) density $N/V_{\rm ref}$ and 
replace the seven Wigner-Seitz cells containing a particle and its six nearest neighbors
by a circle of the same total area.
Such an assumption leads to the value of $R_c^0= (7V_0/\pi)^{1/2} \approx 1.39 a$. 
As we only consider the crystal, in which the density is highly inhomogeneous, however, 
$R_c^0$ is expected to be smaller. As a more appropriate alternative, we extract the value from MC 
simulations of the real-spring system with $V=N V_0$, i.e., the target of our mapping. Specifically, we 
compute the isotropic pair correlation function defined as
\begin{align}
g(r) = \frac{V}{2\pi r N^2}\left\langle
\frac{1}{2} \sum_{i\neq j} \delta (r-r_{ij}) \right\rangle.
\end{align}
By simply probing the distance at which $g(r)$ takes its first minimum,
we find $R_c^0 \approx 1.34a$, compare Fig.~\ref{fig:dft_formulation}.
As the effect from the discontinuity of $u_{\rm pel}(r)$ 
at $R_c^0$ (see, the red solid line in Fig.~\ref{fig:dft_formulation}) 
is minimized with $R_c^0 = 1.34a$, 
the convergence rate turns out to be faster than the minimization with $R_c^0 = 1.39a$.
As the results are hardly affected by this small change of the parameter
and the value $R_c^0=1.34a$ seems to be sufficiently accurate, we use it from now on.
For systems under constant pressure with volumes
other than $V = V_{\rm ref}$ (see Sec.~\ref{sec:elastic}), we use
\begin{align} \label{eq:cutoff_scaling}
R_c = R_c^0\, \sqrt{\frac{V}{V_{\rm ref}}}.
\end{align}
This is the only external input necessary during the formulation of our DFT in Sec.~\ref{sec:DFT}.

The condition (ii) cannot be directly imposed, since the lattice structure is usually not an input 
but the result of a calculation based on a prescribed interaction.
The fact that this condition is not automatically fulfilled even if the real- and pseudo-spring 
systems are at the same density 
is related to another aspect of a pair-potential system in general: the lattice may be imperfect, 
as indicated in Fig.~\ref{fig:pseudo}(c).
The real-spring system with labeled particles assigned to the lattice is completely free from defects, 
whereas the mismatch between the range of the pairwise interaction and the desired lattice structure suggests 
that there should be some vacancies or interstitials.
Moreover, if there were such defects, Eq.~\eqref{eq:Z_pseudo} should also be corrected by additional factors 
addressing the number of possible defect configurations.
Hence, we modify $u_{\rm pel}(r_{ij})$ such that it yields a zero vacancy concentration, $n_{\rm vac}=0$, 
as an equivalent requirement to the above condition of an equal cell structure.
The only way to do so while leaving $R_c$ invariant is to
tune the depth of the pseudo-spring potential in Eq.~\eqref{eq:pseudo_spring} by an offset value $u_{\rm pel}^0$, 
which we understand as follows.
On the one hand, if $\langle u_{\rm pel} \rangle \ll 0$, the total elastic energy is lowered by forming additional contacts with new neighbors. This undesired effect results in the undesired formation of 
interstitials or aggregates. 
On the other hand, when $\langle u_{\rm pel} \rangle \gg 0$, vacancies are generated.
Closing this section on the mapping, we are now ready to formulate our DFT. 
The only remaining mapping parameter $u_{\rm pel}^0$ will be determined within the DFT framework.


%

\subsection{Density functional theory}
\label{sec:DFT}


Following previous studies~\cite{vanTeeffelen2008, Roth2012}, we consider two unit cells of 
a hexagonal lattice in a rectangular base with periodic boundary conditions in $x$ and $y$ directions. 
The volume and the number of particles of the
two unit cells are denoted by $V_{\rm cell}$ and $N_{\rm cell}$, respectively.
Our starting point is to construct a grand canonical free energy functional 
$\Omega ([\rho (\vec{r})])$ 
the value of which at its minimum corresponds to the equilibrium grand potential 
$\Omega_0 (T, \mu, V;m, k_{\rm el}, \eta_0, u_{\rm pel}^0)$ 
in the grand canonical ensemble at fixed temperature $T$, chemical potential $\mu$, and volume $V$.
The free parameters employed here are the magnetic moment $m$, 
the spring constant $k_{\rm el}$, and 
the packing fraction $\eta_0$ of the reference system with $V_{\rm cell}=N_{\rm cell} V_0 = \sqrt{3}a^2$ 
as defined in Sec.~\ref{sec:model}.
In addition, we will complete the theory, by specifying 
the yet to be determined offset for the pseudo-spring potential $u_{\rm pel}^0$.
As we have also employed an additional condition for $n_{\rm vac}$ in the previous section, however, 
each resultant equilibrium density profile corresponds to a parameter set of $(m, k_{\rm el}, \eta_0)$. 


We write 
\begin{align}
\Omega ([\rho (\vec{r})]) = \mathcal{F}([\rho(\vec{r})]) -\mu \int {\rm d}\vec{r} \rho(\vec{r}),
\end{align}
where the Helmholtz free energy functional $\mathcal{F}\equiv \mathcal{F}_{\rm id}([\rho (\vec{r})]) +\mathcal{F}_{\rm exc}([\rho (\vec{r})])$ consists of the ideal gas term and the excess functional,
which read
\begin{align}
&\mathcal{F}_{\rm id}([\rho (\vec{r})]) = \beta^{-1} \int {\rm d}\vec{r} \rho(\vec{r}) [\ln{\{\Lambda^2 \rho(\vec{r})\}}-1], \nn \\
&\mathcal{F}_{\rm exc}([\rho (\vec{r})];m, k_{\rm el}, \eta_0, u_{\rm pel}^0) = 
\mathcal{F}_{\rm m} +\mathcal{F}_{\rm el}+\mathcal{F}_{\rm st},
\end{align}
respectively.
Here, $\Lambda$ denotes the (irrelevant) thermal wave length. 
Regarding the elastic and magnetic interactions, we employ the mean-field functional in the form
\begin{align} \label{eq:mean-field}
\mathcal{F}_{\rm el, m} \equiv 
\frac{1}{2}\int {\rm d}\vec{r}\, {\rm d}\vec{r}\,'\,\rho(\vec{r}) u_{\rm pel, m}(|\vec{r}-\vec{r}\,'|)\rho(\vec{r}\,'),
\end{align}
while we adopt fundamental measure theory~\cite{Roth2012} for $\mathcal{F}_{\rm st}$.
Each excess functional is calculated with the aid of the Fourier convolution theorem.
The detailed forms of the Fourier transforms of $u_{\rm pel}$ and $u_{\rm m}$ used in our calculations 
are described in Appendix~\ref{app:FT_of_energies}.

As in Refs.~\cite{Oettel2010, Roth2012, Lin2018}, we minimize $\Omega$ for a prescribed bulk density 
\begin{align} \label{eq:rho_vac}
\rho_{\rm bulk} \equiv \frac{N_{\rm cell}}{V_{\rm cell}}=\frac{2(1-n_{\rm vac})}{V_{\rm cell}}.
\end{align}
Here, the chemical potential is obtained as an output of the calculation.
Specifically, the minimization process consists of two distinct stages.
At the first stage, we minimize $\mathcal{F}$ for fixed $n_{\rm vac}$  
using the Picard iteration algorithm
\begin{align}
\rho^{(i+1)} (\vec{r}) = \alpha \tilde{\rho}^{(i)}(\vec{r}) +(1-\alpha)\rho^{(i)}(\vec{r}),
\label{eq_PI}
\end{align}
with a mixing parameter $\alpha$~\cite{Roth2010}, 
where
\begin{align}\label{eq:functional_derivative}
\tilde{\rho}^{(i)}(\vec{r})= \frac{1}{\Lambda^2} \exp{\left[ -\beta \frac{\delta \mathcal{F}_{\rm exc}([ \rho^{(i)} (\vec{r})])}{\delta \rho^{(i)} (\vec{r})} +\beta \mu^{(i)} \right]}.
\end{align}
As we have prescribed the bulk density not the chemical potential, 
$\mu^{(i)}$ is updated in each iteration step to keep the average density constant.
Our convergence criterion for minimization at this stage is 
$(\mathcal{F}^{(i)}-\mathcal{F}^{(i+1)})/\mathcal{F}^{(i)} < 10^{-15}$, where $\mathcal{F}^{(i)} = \mathcal{F}([\rho^{(i)}(\vec{r})])$.
Then we determine the equilibrium density profile by further minimizing $\mathcal{F}/N$ 
with respect to $n_{\rm vac}$. 
In practice, we vary $n_{\rm vac}$ by controlling $V_{\rm cell}$ while $\rho_{\rm bulk}$ is fixed.
Comparing the values of $\mathcal{F}/N$ obtained for each $n_{\rm vac}$, 
we determine the vacancy concentration $n_{\rm vac}$ and the equilibrium density profile $\rho(\vec{r})$
with which the free energy per particle $\mathcal{F}/N$ is minimized.
These procedures are repeated until we have an accuracy of $10^{-6}$ for $n_{\rm vac}$.
In the calculations, we set $\Lambda = 1$.

Now, we determine the values of $u_{\rm pel}^0$ to close our DFT. 
Specifically, we first perform the two-stage DFT minimization as discussed above
to find corresponding vacancy concentrations for given values of $u_{\rm pel}^0$. 
Then we choose the value of $u_{\rm pel}^0$ for which the vacancy concentration is zero, i.e., $n_{\rm vac} = 0$.
As described in Eq.~\eqref{eq:rho_vac}, 
controlling $n_{\rm vac}$ in DFT calculations involves the change in $V_{\rm cell}$. 
Therefore, we also use a rescaled value of $R_c$ given by Eq.~\eqref{eq:cutoff_scaling} 
when $n_{\rm vac}\neq 0$.

\begin{figure}
\includegraphics[width=8.6cm]{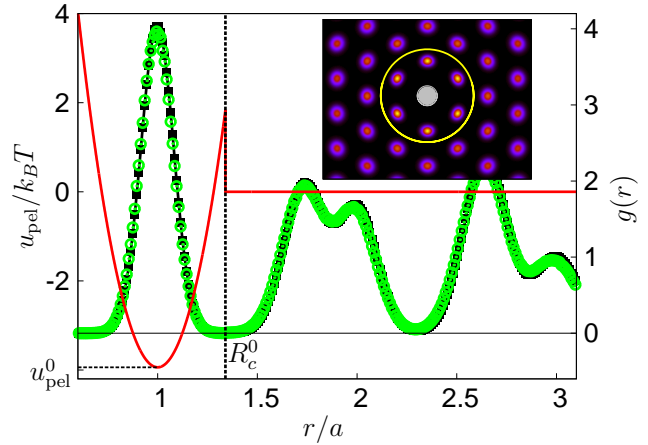}
\caption{\label{fig:dft_formulation}
Formulation of a DFT for the bead-spring model.
We compare pair correlation functions $g(r)$ from the real and pseudo-spring MC simulations of the reference system
with $V=V_{\rm ref}$,
which are depicted by black and green symbols, respectively.
First, black square symbols show $g(r)$ obtained from the MC simulation of the real-spring system.
As depicted by the black vertical line, the distance at the minimum point between the first and the second peak  
in $g(r)$ is chosen as $R_c^0$. 
Here, $ka^2/k_B T=100$, $\eta_0 = 0.3$, $m/m_0 =0$ and $N=480$ are used and we find $R_c^0= 1.34$.
In the inset, the anisotropic pair correlation function $g(\vec{r})$ obtained from the real spring 
MC-simulation results is also displayed, together with the yellow line indicating $R_c^0$.
As shown, the yellow line is far enough from the neighboring peaks and we confirm that the isotropic
cutting off of the springs is a reasonable approach.
The pair correlation function $g(r)$ obtained from pseudo-spring systems 
with $u_{\rm pel}^0 =3.94k_B T$ is represented by green circle symbols. 
The agreement between the real- and pseudo-spring simulations is manifested clearly. 
Finally, the red solid line presents the elastic energy of pseudo-springs $u_{\rm pel} (r)$ 
defined in Eq.~\eqref{eq:pseudo_spring}. 
}
\end{figure}

Performing first the minimization with $m=0$ and without a hard core repulsion ($\eta_0=0$), 
we confirm that the pseudo-spring potential indeed admits a freezing transition,
which is the prerequisite of the mapping.
Then we perform MC simulations with $\tilde{\mathcal{H}}_{\rm pel}$, together with $\mathcal{H}_{\rm st}$,
to verify the mapping at finite packing fraction, 
using the value of $u_{\rm pel}^0$ obtained from such a DFT as an input.
As manifested in Fig.~\ref{fig:dft_formulation}, 
the agreement between the real- and the pseudo-spring systems is quantitatively excellent.
Therefore, we conclude that the DFT approach to systems with distinguishable particles is
successfully formulated with the pseudo-spring potential between indistinguishable particles
and ready to use in the presence of magnetic interactions as a model for ferrogels.

\section{Elastic properties of ferrogels}
\label{sec:elastic}
%
We finally demonstrate the utility of the theory to investigate the mechanical properties of ferrogels, 
varying the magnetic moment and the density of the particles.
In particular, we probe the system at constant pressure $p = k_B T/a^2$
and determine the volume $V$ and the responses to elastic deformations $\vec{\nabla} \vec{u}$,
where $u$ is the displacement field. 
With the component of the corresponding linear strain tensor~\cite{Landau1986}
\begin{align}
\epsilon_{ij} =\frac{1}{2}(\nabla_i u_j + \nabla_j u_i),
\end{align}
the stress tensor and the stiffness tensor are defined as
\begin{align} \label{eq:def_modulus}
\sigma_{ij} = \frac{\partial f}{\partial \epsilon_{ij}}, \quad 
C_{ijkl} = \frac{\partial \sigma_{ij}}{\partial \epsilon_{kl}},
\end{align}
where $f \equiv F/V$ is the density of the Helmholtz free energy $F$. 
For two-dimensional hexagonal lattices, the stiffness tensor 
\begin{align}
C_{ijkl} = K\delta_{ij}\delta_{kl} +G (\delta_{ik}\delta_{jl} +\delta_{il}\delta_{jk} -\delta_{ij}\delta_{kl}).
\end{align}
can be expressed in terms of only two independent elastic constants,
namely the bulk modulus $K$ and the shear modulus $G$ because of the symmetry~\cite{Chaikin2000}.

In the following, we describe in Sec.~\ref{sec:impl} how to calculate the volume and elastic constants 
in MC simulations for the real- and pseudo-spring systems and using pseudo-potentials from our DFT treatment
we compare the results of those methods in Sec.~\ref{sec_res2}.
Henceforth, $V$ and the elastic constants, namely $K$ and $G$, are measured 
in units of $a^2$ and $k_B T/a^2$, respectively.


\subsection{Implementation} 
\label{sec:impl}

\subsubsection{MC simulation}
In the case of MC simulations,
we employ the isobaric-isothermal ($TpN$) ensemble~\cite{Wood1968, Eppenga1984}
and vary the rectangular lengths $L_x$ and $L_y$ independently~\cite{Boal1993}.
Accordingly, random walks in terms of $\ln{L_x}$ and $\ln{L_y}$ have been performed 
with the detailed balance condition~\cite{Frenkel2002}
\begin{align}
\frac{\omega(V\to V')}{\omega(V'\to V)}
&=\exp \left\{ -\beta[U(\vec{s}\,'^N, V')-U(\vec{s}^N,V) \right. \nn \\ 
 	& \left.  +p(V'-V)] + (N+1)\ln{(V'/V)} \right\},
\end{align} 
where $\omega$ is the transition rate corresponding to the volume changes 
and $\vec{s}^N=\{ \vec{s}_i \}$ are the scaled coordinates
defined by $\vec{s}_i \equiv V^{-1/2} \vec{r}_i$ for $i=1,\ldots, N$.
We then compute the volumes of the systems, simply taking the average of $V \equiv \langle L_x L_y \rangle$ and extract $K$ and $G$ from the fluctuations $\langle (\Delta V)^2 \rangle$, 
$\langle (\Delta L_x)^2 \rangle$, and $\langle (\Delta L_y)^2 \rangle$~\cite{Boal1993}.
The elastic constants are computed from the real-spring as well as the pseudo-spring MC simulations 
to verify the validity of the mapping. In particular, for the pseudo-spring MC simulation,
we use the values of the offset $u_{\rm pel}^0$ obtained from the density functional calculations as inputs, 
while the average volumes extracted from the corresponding real-spring MC 
simulations are employed to determine the value of the cut-off $R_c$,
with the aid of Eq.~\eqref{eq:cutoff_scaling}.

\subsubsection{DFT calculation}

Computation of thermodynamic quantities under the given pressure in DFT
is formally not straight-forward, because the theory is based on the grand-canonical ensemble.
In a finite system, for example, the structure depends significantly on the specific choice
 of the statistical ensemble~\cite{Gonzalez1997, Gonzalez2004, delasHeras2016}.
%
Due to the equivalence of the ensembles in the thermodynamic limit, however, 
one can still regard the density profiles obtained from DFT also as minima in the isobaric ensemble
as long as large systems are considered.
Hence, the requirement to compare the DFT results to MC simulations at constant pressure
does not represent a conceptual problem for our study.
Specifically, we first compute pressures at various volumes $V_{\rm cell}$ from the relation $p=-\Omega_0/V_{\rm cell}$ 
and choose the volume at which the pressure reaches the prescribed value.
The accuracy of the volume is $10^{-5}(2V_0)$ and the vacancy concentration is fixed during 
the procedures by fine-tuning $u_{\rm el}^0$ with an accuracy of $10^{-4}k_B T$.
The detailed procedures are exemplified in Appendix~\ref{app:DFT_NpT}.


Since we have direct access to the free energy, 
we can directly compute the elastic constants deforming the system.
We consider four types of deformations as follows:
\begin{align}
{\tensor{\epsilon}} =  
	\left( {\begin{array}{cc} \epsilon_K & 0 \\ 0 & \epsilon_K \\ \end{array} } \right), 
	\left( {\begin{array}{cc} \epsilon_x & 0 \\ 0 & 0 \\ \end{array} } \right), 
	\left( {\begin{array}{cc} 0 & 0 \\ 0 & \epsilon_y \\ \end{array} } \right), 
	\left( {\begin{array}{cc} \epsilon_G & 0 \\ 0 & -\epsilon_G \\ \end{array} } \right). 
\end{align}
Specifically, we first deform the system according to the given strain, simultaneously 
controlling the density. Then we obtain the equilibrium density profile of the deformed system
adjusting the vacancy concentration equivalently to the undeformed system as described in Sec.~\ref{sec:DFT}.
We note that two types of deformation are enough to determine the two unknown $K$ and $G$.
Examining four types of deformation, we verify the consistency of the theory.
However, the deformations involving a volume difference seem to involve inelastic changes:
shifts of the vacancy concentration in the equilibrium step indeed imply changes
in the number of particles in the unit-cell, which would not occur during genuinely elastic deformations.
To minimize such inelastic contributions, we utilize the identity, $p \equiv -\Omega_0 /V_{\rm cell}$,
which does not involve any deformations, 
instead of directly computing the bulk modulus $K$ from the definition of Eq.~\eqref{eq:def_modulus}.
Specifically, the elastic constants are calculated from
\begin{align} \label{eq:K_comp}
K&=-\frac{\{p(\epsilon_K)-p(0)\}+\{p(0)-p(-\epsilon_K)\}}{4\epsilon_K}, \\ \label{eq:G_comp}
G&=\frac{f(\epsilon_G)+f(-\epsilon_G)-2f(0)}{4{\epsilon_G}^2}.
\end{align} 
Here the pressure of the deformed systems is computed from $p(\epsilon_K) = -\Omega_0 (\epsilon_K)/V_{\rm cell}(\epsilon_K)$, 
where $\Omega_0 (\epsilon_K)$ is the grand potential at which the density functional is minimized 
under the given constraint due to a deformation.
We note that, in every case, the differences in the values of $K$ are less than 10\%, compared to the results 
for $K$ that we have calculated in analogy to Eq.~\eqref{eq:G_comp} via the changes in $f$, instead of using 
the pressure as in Eq.~\eqref{eq:K_comp}.


Before we proceed, let us make a few technical remarks.
First, the minimization of the functionals at $n_{\rm vac}=0$ requires large computational resources
since a small value of the Picard iteration parameter $\alpha$ in Eq.~\eqref{eq_PI} should be used 
to guarantee the convergence of the minimization. 
To calculate the equilibrium density in a reasonable time scale, 
we chose to fix the iteration parameter at $\alpha=0.001$ in general, 
at the cost of loosening the strict condition to impose a zero vacancy concentration.
Values of $\alpha$ smaller than that are only used for a few cases in which the minimization eventually fails.
With this constraint, we could minimize the free-energy functional 
up to $n_{\rm vac}= 0.0006 \pm 1.0\times 10^{-6}$ for $\eta=0.3$ and $0.5$, 
and $n_{\rm vac}=0.015\pm 1.0\times 10^{-6}$ for $\eta=0.8$. 
Secondly, as is well known, the numerical treatment of long-range interactions with periodic boundary condition
would require sophisticated techniques~\cite{Allen1989, Arnold2005}. In our unit-cell DFT calculations,
we obviously neglect the Fourier components the wave lengths of which are longer than the unit-cell size.
To bypass such complications and provide a fair comparison, 
with the MC simulations we simply cut the magnetic interaction beyond the nearest-neighbor interaction
as an approximation~\cite{Goh2018, Ider2019}.


\subsection{Results}
\label{sec_res2}

\begin{figure}
\includegraphics[width=8.6cm]{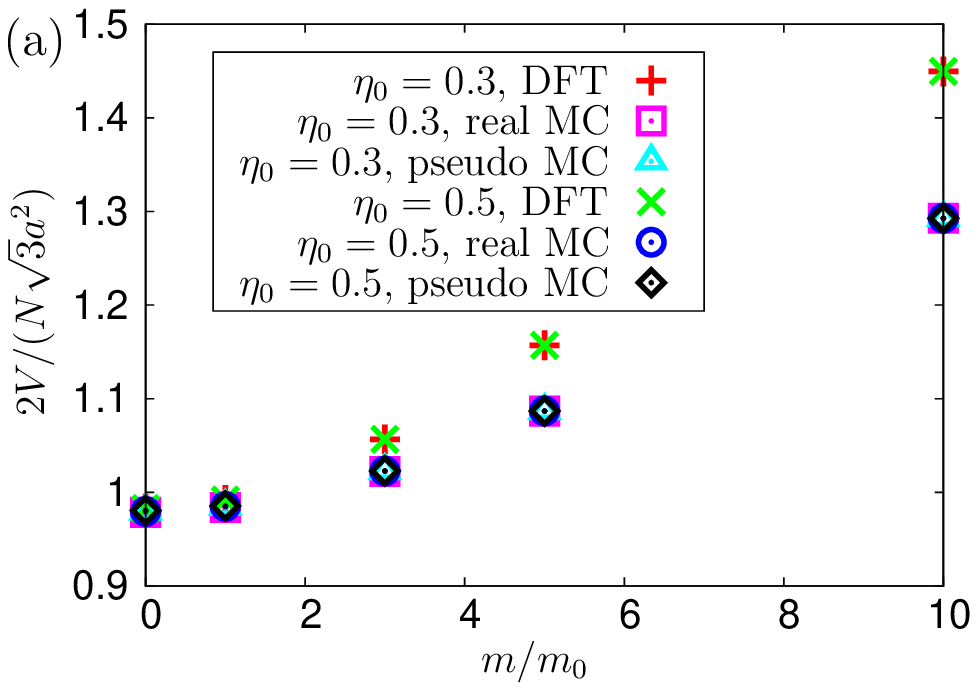}
\includegraphics[width=8.6cm]{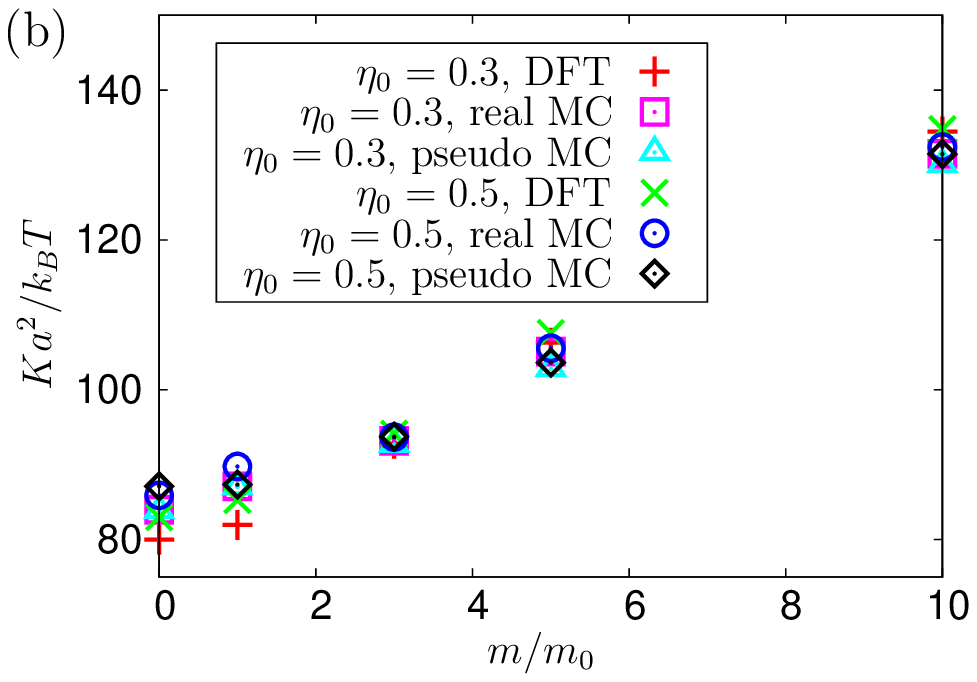}
\includegraphics[width=8.6cm]{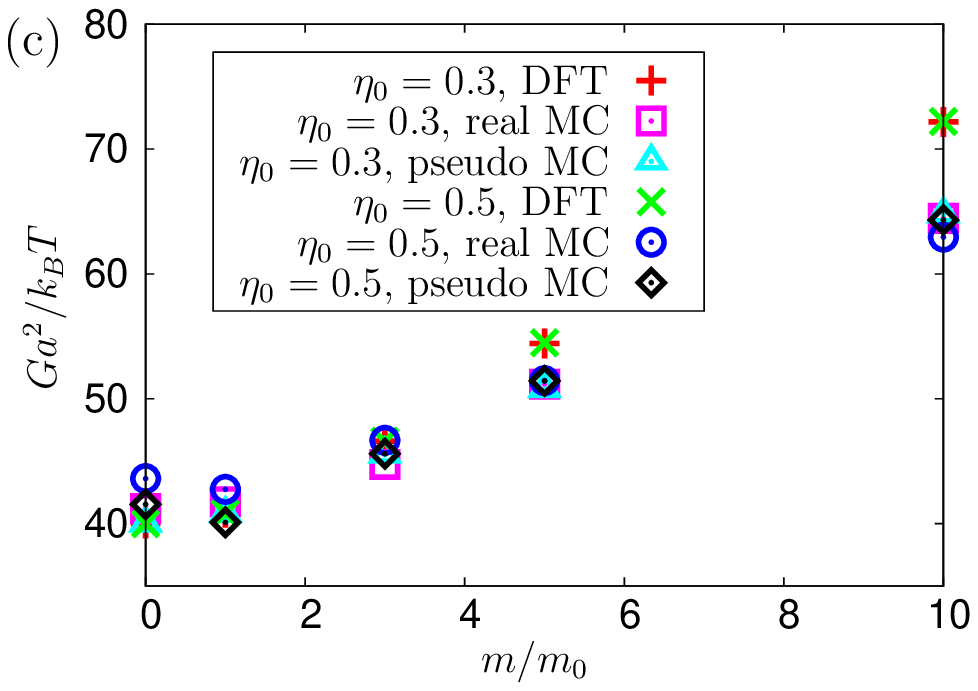}
\caption{\label{fig:result_low} 
For systems of low packing fractions, i.e., $\eta_0 = 0.3$ and $0.5$,
(a) the volume $V$, (b) the bulk modulus $K$, and (c) the shear modulus $G$ are presented as functions of
the repulsive magnetic dipole moment $m$.
Apparently, the DFT overestimates the volume $V$ for elevated values of $m$.
Rough agreement among the theory and the MC simulations of both the real- and pseudo-spring systems are observed 
for the both elastic constants $K$ and $G$. 
Here, $\epsilon_K = 0.00025$, $\epsilon_G = 0.00025$ and $N=480$ have been used.
}
\end{figure}

We explore the following sets of parameters: $(k_{\rm el}, \eta_0)=(100.0, 0.3), (100, 0.5), (100.0, 0.8)$, 
each for several magnetic moments $m\leq12$.
Again, we note here that $\eta_0$ corresponds to the packing fraction of the reference system with $V=V_{\rm ref}$,
conveying the information of the diameter of particles. 
The conventional packing fraction is denoted by $\eta \equiv N/V$ and is not a fixed variable 
as we consider the isobaric ensemble, see the paragraph below Eq.~\eqref{eq_Hst} for details. 
In Fig.~\ref{fig:result_low}, we first present the results at low packing fractions, 
i.e., $\eta_0 = 0.3$ and $0.5$, 
for which the conventional hard disk system is in the fluid phase~\cite{Mak2006,Roth2012,Engel2013}. 
Therefore, the crystallization in this low packing fraction regime is due to the elastic interaction.
As one can see, the DFT and the real- and pseudo-spring MC simulations agree well 
with each other, except that the DFT overestimates the volumes,
especially for large values of $m$. 
As we adopt the mean-field functionals, 
the repulsion effects from the energy seem to be exaggerated, leading to the increases in volumes.
Overall, the volume $V$, the bulk modulus $K$, and the shear modulus $G$ increase 
as the magnetic moment $m$ increases.

\begin{figure}
\includegraphics[width=8.6cm]{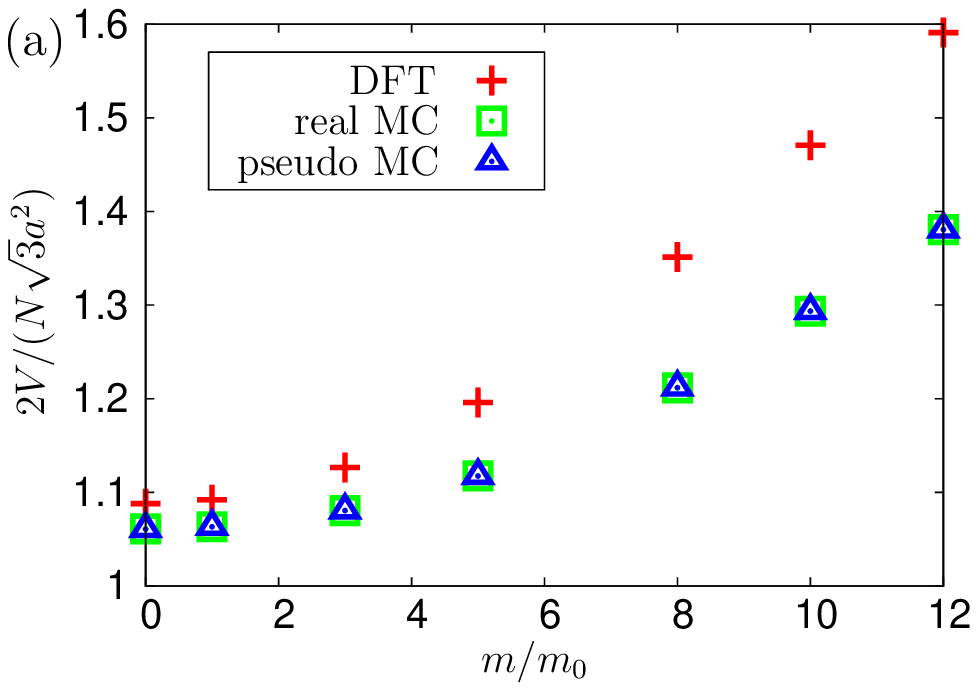}
\includegraphics[width=8.6cm]{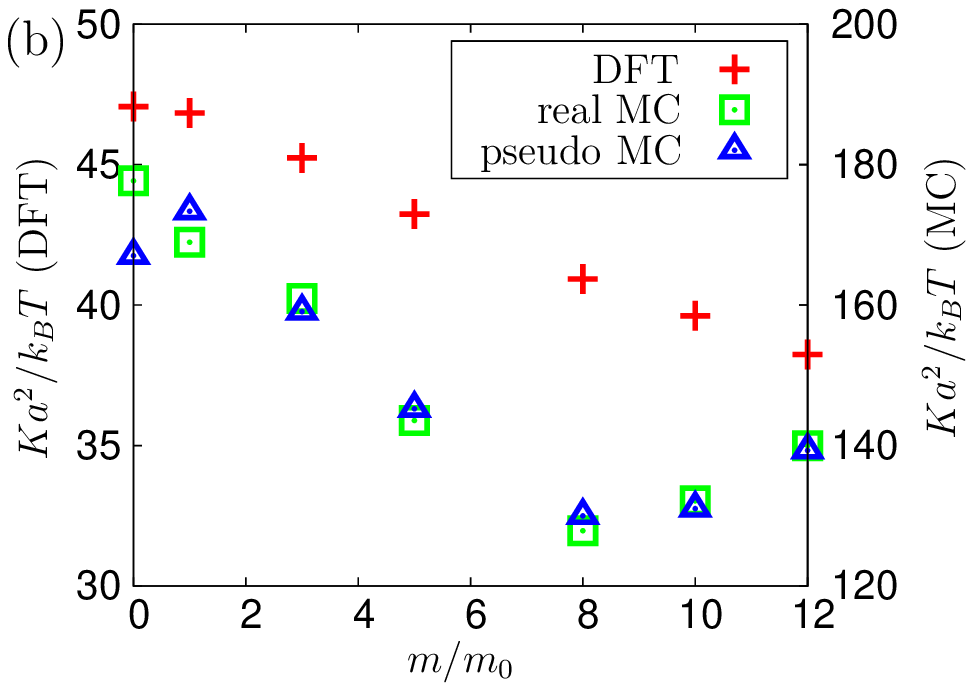}
\includegraphics[width=8.6cm]{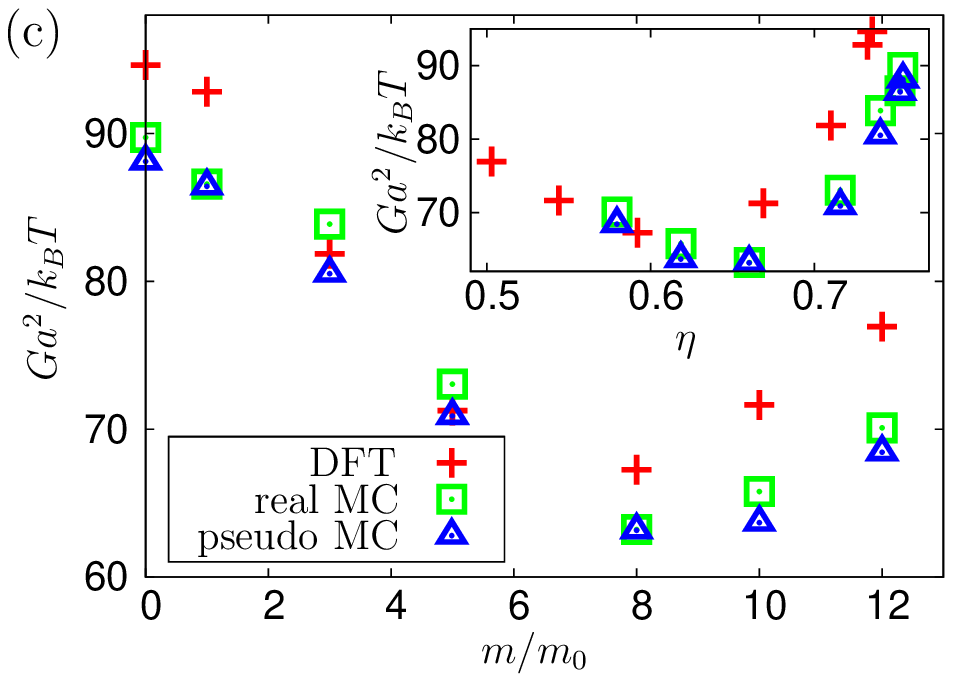}
\caption{\label{fig:result_high} 
For systems with a high packing fraction, $\eta_0 = 0.8$,
(a) the volume $V$, (b) the bulk modulus $K$, and (c) the shear modulus $G$ are presented 
as functions of the magnetic dipole moment $m$. 
As in Fig.~\ref{fig:result_low}, the DFT overestimates the volume $V$.
For the elastic constants, 
the agreement among the theory, the real MC simulations and the pseudo MC simulations is again quite reasonable 
for the shear modulus $G$. However, the DFT significantly underestimates 
the bulk modulus $K$. The reason for this large deviation is very likely
a rather large value of the vacancy concentration, 
$n_{\rm vac} =0.015$, see the main text for details. In the inset of (c),
the values of the shear modulus are additionally presented 
as a function of the packing fraction $\eta \equiv N/V$.
Here, $\epsilon_K = 0.00025$, $\epsilon_G = 0.00025$ and $N=120$ have been used.
}
\end{figure}

Meanwhile, results with $\eta_0 =0.8$ are shown in Fig.~\ref{fig:result_high}. 
In this case, we expect strong contributions from the steric forces.
We first note that a significant deviation of the theory from the MC simulations in the bulk modulus 
is observed. Quantitatively, the bulk moduli obtained from the DFT are 
approximately one fourth of those obtained from the MC simulations.
This is mostly due to the fact that we had to use a quite large value, i.e., $n_{\rm vac}\approx 0.015$,
for the vacancy concentration because of the technical reasons related to the computational time and 
the consequent choice of the mixing parameter $\alpha$ (see Sec.~\ref{sec:impl}). 
Indeed, using the same vacancy concentration, 
we also obtain similar deviations in $K$ at $\eta_0=0.3$ or $0.5$.
Moreover, a good agreement between the real- and pseudo-spring MC simulations is still observed, 
confirming that such a deviation of the DFT does not indicate a failure of our mapping. 
Surprisingly, the behavior of $G$ is still well predicted by the DFT.
In contrast to $K$, computation of $G$ does not involve any changes in volume and consequently 
both the density and the vacancy concentration remain approximately constant during deformation. 
(In contrast to that, deformations involving volume changes are always accompanied by shifts 
of the vacancy concentration in equilibrium.)
Therefore, we speculate that the vacancy concentration plays a much smaller role 
in the case of the volume-conserving deformation in DFT, for which $n_{\rm vac}$ remains basically the same. 
To conclude, except for the technical issue discussed, the DFT provides qualitatively correct trends 
even at high packing fractions.

Remarkably, we observe new response behaviors of $K$ and $G$ to an increase in $m$
for the high packing fraction, which are opposite to the low packing fraction. 
For $\eta_0 = 0.8$, $K$ and $G$ decrease as $m$ increases, while $V$ is still an increasing function of $m$.
Emergence of such two different scenarios originates from the composite nature of the ferrogels:
here, we are observing different types of crossovers 
from the hard-disk crystals (high packing fraction, small $m$) 
or the harmonic crystals (low packing fraction, small $m$)
to the elastic-dipolar crystals (low packing fraction, large $m$).

To understand the phenomena in detail, we first note that $V$ always increases as $m$ increases
because the magnetic interaction is repulsive.
Then for low packing fractions, the magnetic repulsion simply causes additional increases 
in the bulk and shear moduli on top of the harmonic crystals. 
In contrast to that, for high packing fractions, the steric force dominates the mechanical properties 
and the system is very stiff (hard-disk crystals) with relatively large $K$ and $G$ for small values of $m$. 
As the magnetic moment $m$ and consequently the volume $V$ increase, 
the packing fraction $\eta$ decreases 
and the contributions from the steric repulsion become insignificant
compared to the dipolar repulsion at some values of the packing fraction around $0.6 \lesssim \eta \lesssim 0.7$
[see, e.g., the inset of Fig.~\ref{fig:result_high}(c)].
These values are slightly smaller than the fluid-crystal coexisting packing fractions of 
the hard-disk systems, 
which are in between 0.68 and 0.73~\cite{Bernard2011, Engel2013, Roth2012, Thorneywork2017, Lin2018}.
Once $\eta$ has decreased enough, as in the low packing fraction regime,
the signature of the elastic-dipolar crystals should be recovered, the elastic properties of which are 
governed by the combination of the spring and dipolar interactions.
Indeed for $m=10$, the volume
and the elastic constants (compare Figs.~\ref{fig:result_low} and~\ref{fig:result_high})
 are quantitatively similar 
among $\eta_0$=0.3, 0.5, and 0.8. As expected, the values of $K$ and $G$ for $\eta_0=0.8$ increase in 
the regime of large magnetic moments, i.e., for $m \gtrsim 8$ where $\eta < 0.70$ 
($\eta \approx$ 0.68 and 0.60 for $m=$5 and 8, respectively).

\section{Summary and outlook}
\label{sec:conclusion}
In this study, we have formulated a DFT for a two-dimensional ferrogel model. 
We have replaced the labeled particles in a state of strictly permanently connected neighboring particles 
by the unlabeled particles in 
a fluid-like state to map the elastic part of the associated energy 
onto a pseudo-potential invariant under permutations.
In particular, we have shown that the mapping provides a plausible approximation for the considered systems and 
their response to magnetic interactions, even though the mapping still leads to some deviations in the
calculated response of the systems. These deviations
have been minimized by fine-tuning the mapping parameters.
Lastly, it has been demonstrated that the elastic properties of ferrogels can be successfully investigated 
in this framework and two scenarios have been identified for the response mechanism of the dipole-spring
system, depending on the packing fraction.
Notably, our DFT approach may also provide a clue for the scale-bridging~\cite{Menzel2014}
between mesoscopic dipole-spring models and the macroscopic description of ferrogels~\cite{Jarkova2003,Bohlius2004}.

A strong feature of the presented DFT approach relying on the pseudo-spring model in two dimensions is 
that it works well independently of the density.
First of all, the mapping is generally well justified even in the low-density regime.
Moreover, the mean-field DFT, which we employed here, provides a good approximation even in the high-density regime 
where we found good agreement between our theoretical treatment and MC simulations, 
especially for the shear modulus. 
In contrast to that, for a one-dimensional model, the prediction of unphysical freezing within mean-field 
DFT restricts the parameter space to intermediate densities and weaker elasticities~\cite{Cremer2017}, where
the pseudo-spring approximation 
itself is less accurate due to large possible gaps between neighboring particles.
We stress that, in general, mean-field approximations become more accurate 
with increasing dimensionality. 
Besides, true long-range order generally exists in three dimensions.
Therefore, we expect our DFT approach to work even better in three dimensions.

Regarding possible experiments to test our theoretically predicted scenarios, 
the high packing fraction of $\eta_0 = 0.8$ needs to be discussed. 
As mentioned in Ref.~\cite{Huang2016}, for certain systems there exists a region of increased 
mechanical stiffness around the magnetic particles. Effectively, this could be equivalent to an increased 
size of the particles. In addition, we note that in three dimensions, the fluid-crystal coexisting densities
are reported as low as $\eta \approx 0.5$~\cite{Hoover1968, Roth2002}.
Indeed, volume concentrations of approximately 50\% have already been reported 
in three-dimensional samples of ferrogels~\cite{Liu2008}, 
implying that the steric force should be taken into account
explicitly. Therefore, decreasing elastic moduli in response to an increase of the 
magnitude of the magnetic moments might be observed in real three-dimensional systems. 
Furthermore, the pressure value employed in this study seems to be small: 
for instance, $p = k_B T/a^2 \sim 2\times 10^{-7}$\,Pa for the system studied in Ref.~\cite{Messing2011} with $a\approx 150$\, nm 
at room temperature. 
Therefore, for direct comparison with experiments, a broad range of pressures 
should be explored in future studies with more realistic settings.

There are several remaining issues. 
First, aligning the magnetic dipoles within the two-dimensional plane and the resulting in-plane anisotropy will
give rise to additional phenomena not observed here for perpendicular dipoles.
To list a few, the aspect ratio of width to height of the unit cell 
will vary depending on the in-plane orientation of the magnetic dipole moments, 
the volume may change 
and the magnetic particles could touch each other, leading to an abrupt change in elastic moduli
even at a relatively low packing fraction~\cite{Annunziata2013}.
As mentioned above, an extension to three dimensions would even strengthen the correspondence to real materials.
Additionally, the mean-field functionals can be replaced by more sophisticated ones, 
especially to consider the long-range nature of the magnetic interaction~\cite{vanTeeffelen2008}.
Apart from that, the response dynamics to external magnetic fields represents another topic of interest. 
Dynamical DFT~\cite{Marconi1999, Archer2004, Stopper2018} 
should be the obvious candidate to study these phenomena in the present context.
\section*{Acknowledgments}
We thank Urs Zimmermann and Christian Hoell for providing numerical codes that were useful for the initiation of this study. 
We also thank Shang-Chun Lin, Martin Oettel, Benno Liebchen, Mehrdad Golkia, Saswati Ganguly, and Robert Evans
for helpful discussions.
This work was supported by funding from the Alexander von Humboldt-Stiftung (S.G.) and
from the Deutsche Forschungsgemeinschaft through the SPP 1681,
grant nos. ME 3571/3 (A.M.M) and LO 418/16 (H.L.).

\appendix

\section{Fourier transform}
\label{app:FT_of_energies}
The functional derivatives of the mean-field functionals in Eq.~\eqref{eq:mean-field}
for the calculation of Eq.~\eqref{eq:functional_derivative} read
\begin{align}
\frac{\delta \mathcal{F}_{\rm el, m}}{\delta \rho(\vec{r})} = &\int {\rm d} \vec{r}\,'\rho(\vec{r}\,')u_{\rm pel,m}(|\vec{r}-\vec{r}\,'|),
\end{align}
which are numerically implemented with the aid of the convolution theorem. 
First, the Fourier transform of the elastic energy $\tilde{u}_{\rm pel}$ can be computed as follows:
\begin{align}
\tilde{u}_{\rm pel} (\vec{k}) =& \int_0^{2\pi} {\rm d}\theta \int_\sigma^{R_c} {\rm d}r \,
r\left\{\frac{1}{2}k_{\rm el} (r-a)^2 -u_{\rm el}^0 \right\} \, e^{-i\vec{k}\cdot \vec{r}} \nn \\
=&\int_\sigma^{R_c}{\rm d}r\,2\pi r J_0(kr) \left\{ \frac{1}{2} k_{\rm el} (r-a)^2 -u_{\rm el}^0 \right\} \nn\\
=&\frac{\pi k_{\rm el}}{k^3} \left[ r J_1 (kr) \left\{ -4 +k^2 (r-a)^2 +ak\pi H_0 (kr) \right\} \right. \nn \\
	&\hspace{1cm}\left. + rJ_0(kr)\left\{ 2kr-ak \pi H_1 (kr) \right\} \right]_\sigma^{R_c} \nn \\
	&-\frac{\pi u_{\rm el}^0}{k} \left[ 2r J_1 (kr) \right]_\sigma^{R_c}.
\end{align}
For the magnetic interaction, 
the Fourier transformation can be performed as follows:
\begin{align}
\tilde{u}_{\rm m}  (\vec{k}) =&\frac{\mu_0 m^2}{4\pi} \int_\sigma^\infty {\rm d}r  \int_0^{2\pi}{\rm d}\theta
	\frac{1}{r^2} e^{-i \vec{k} \cdot \vec{r}} \nn \\
=&\frac{\mu_0 m^2}{2} \int_\sigma^\infty {\rm d}r \frac{J_0 (kr)}{r^2} \nn \\
=&\frac{\mu_0 m^2}{2} \left[ 
	-\frac{1}{r} {}_1{}F_2 \left(-\frac{1}{2};\frac{1}{2},1;-\frac{1}{4}k^2 r^2 \right) \right]_\sigma^\infty, 
\end{align}
where ${}_p{}F_q$ is the generalized hypergeometric function.
For $r \to \infty$,
\begin{align}
{}_1{}F_2 & \left(-\frac{1}{2};\frac{1}{2},1;-\frac{1}{4}k^2 r^2 \right) \nn \\
	&= kr + \frac{\cos{kr}-\sin{kr}}{\sqrt{\pi}k^{3/2}} r^{-3/2} +O\left( r^{-5/2} \right).
\end{align}
We finally obtain
\begin{align}
&\tilde{u}_{\rm m} (\vec{k}) = \frac{\mu_0 m^2}{2} \left[ 
	-k +\frac{1}{\sigma}{}_1{}F_2 \left(-\frac{1}{2};\frac{1}{2},1;-\frac{k^2 \sigma^2}{4} \right)
		\right].
\end{align}
The generalized hypergeometric functions were implemented using the Arb library~\cite{Johansson2017}.

\section{DFT minimization}
\label{app:DFT_NpT}
In this section, we describe the procedure of computing the equilibrium profile at prescribed pressure
within the grand-canonical DFT,
which is exemplified in Table~\ref{table:DFT}.
First, the volume $V$ and the offset $u_{\rm el}^0$ 
are fixed (the first and the second column in Table~\ref{table:DFT}. 
We then compute the free energy per particle $\mathcal{F}/N_{\rm cell}$, 
varying the vacancy concentration $n_{\rm vac}$ (the third column).
At this stage, we control $n_{\rm vac}$ by changing $V_{\rm cell}$ while fixing the density $\rho_{\rm cell}$,
see  Eq.~\eqref{eq:rho_vac}.
We choose the value of $u_{\rm el}^0$ where $\mathcal{F}/N_{\rm cell}$ is minimized 
at the prescribed value of the vacancy concentration 
(for the parameter set in Table~\ref{table:DFT}, $n_{\rm vac}=0.0006$).
At this stage, the pressure corresponding to the fixed volume is determined simultaneously (the fourth column).
Now, we change the volume, 
varying the density $\rho_{\rm cell}$ while fixing the number of particles $N_{\rm cell}$,
and repeat the above procedure to calculate corresponding pressures.
Finally, we obtain the volume and corresponding density profile,
comparing pressures with the prescribed pressure value, i.e., $pa^2/k_B T = 1$.

\begin{table}
\begin{tabular}{c|c|c|c}
\hline
\hline
$V_{\rm cell}/2V_0$ & $u_{\rm el}^0/k_B T$ & $n_{\rm vac}$ & $pa^2/k_B T$ \\
\hline
 $\ \ $0.98250$\ \ $ & $\ \ $2.7419$\ \ $ & $\ \ $0.0006002$\ \ $ & $\ \ $$\ \ $ \\ 
 $\ \ $$\ \ $ & $\ \ $2.7420$\ \ $ & $\ \ $0.0006000$\ \ $ & $\ \ $1.000609$\ \ $ \\ 
 $\ $$\ $ & $\ $2.7421$\ $ & $\ $0.0005999$\ $ & $\ $$\ $ \\
\hline
 $\ $0.98251$\ $ & $\ $2.7422$\ $ & $\ $0.0006002$\ $ & $\ $$\ $ \\
 $\ $$\ $ & $\ $2.7423$\ $ & $\ $0.0006000$\ $ & $\ $0.999613$\ $ \\
 $\ $$\ $ & $\ $2.7424$\ $ & $\ $0.0005998$\ $ & $\ $$\ $ \\
\hline
\hline
\end{tabular}
\caption{\label{table:DFT}An example of the DFT computation procedure. 
Here, $ka^2 /k_B T=100$, $\eta_0 = 0.3$, $R_c^0/a=1.34$.}
\end{table}

\bibliography{DFT_2D_ferrogel}

\end{document}